\documentclass[lettersize,journal]{IEEEtran}
\usepackage{amsmath,amsfonts}
\usepackage{algorithmic}
\usepackage{algorithm}
\usepackage{array}
\usepackage[caption=false,font=normalsize,labelfont=sf,textfont=sf]{subfig}
\usepackage{textcomp}
\usepackage{stfloats}
\usepackage{url}
\usepackage{verbatim}
\usepackage{graphicx}
\graphicspath{{eps/}}
\usepackage{cite}
\begin{document}

\title{Multi-Octave Interference Detectors with Sub-Microsecond Response}

\author{Mohammad~Abu~Khater,~\IEEEmembership{senior member,~IEEE,}, and Dimitrios~Peroulis,~\IEEEmembership{fellow,~IEEE,}

        % <-this % stops a space
\thanks{The authors are with Purdue University, school of Electrical and Computer Engineering. Contact: mabukhater@ieee.org}% <-this % stops a space
\thanks{Manuscript received April 19, 2021; revised August 16, 2021.}}

\markboth{TMTT,~Vol.~14, No.~8, August~2021}%
{Shell \MakeLowercase{\textit{et al.}}: A Sample Article Using IEEEtran.cls for IEEE Journals}


\maketitle

\begin{abstract}
High-power interferers are one of the main hurdles in wideband communication channels. To that end, this paper presents a wideband interferer detection method. The presented technique operates by sampling the incoming signal as an input, and produces the frequency and the power readings of the detected interferer. The detection method relies on driving an open circuit stub, where the voltage is proportional to the power of the interferer, and the standing wave pattern is an indicator of its frequency. This approach is feasible over multi-octave bandwidth with a wide power dynamic range. The concept is analyzed for design and optimization, and a prototype is built for a proof-of-concept. The measured results demonstrate the ability to detect an interferer within the 1--16 GHz frequency range, with a power dynamic range between -20 to 20 dBm. The detection concept is also fitted with different types of tunable bandstop filters (BSFs) for automatic detection and suppression of the interferer if its power exceeds a programmable threshold. With a measured response time of 500 ns, the presented method is a technology enabler for wideband receivers. 
\end{abstract}


\section{Introduction}

\IEEEPARstart{W}{ideband} receivers are essential for numerous applications such as opportunistic radios and high-throughput, multi-channel links. Exposing receiver circuits to a very wide spectrum, however, makes them vulnerable to strong interferers. This can result in desensitization, deteriorated linearity performance, and possibly damaging the receiver. Consequently, a diverse amount of technologies and techniques are researched to detect and suppress interferers in wideband systems. 

Signal processing techniques, analog and digital, have been implemented to detect and suppress interferers for various wideband applications such as UWB \cite{uwb_digital_suppression,gnss_digital_interf_detec,dsp_beamforming,mixer_detection}. Such techniques benefit from the agility of digital systems, or the speed of analog circuits. To operate properly, however, those techniques rely on the linearity, and the power reliability, of the RF front-end, which could be a limitation if a high-power interferer exists. The operating bandwidth is also typically limited.

RF power limiters have shown a relatively fast response (nanoseconds in \cite{abbas_resonator}) while protecting the rest of the RF chain. This is favorable in pulsed interferers, where it is necessary to attenuate a short high-energy pulse. Wideband attenuators are available in various technologies such as plasma \cite{abbas_resonator}, MEMS \cite{mems_self_actuating_limiter} or solid-state \cite{pin_limiter,gan_robust}. The use of wideband attenuators results in an adverse suppression of the whole operating band, suppressing the signal of interest. Alternatively, widely tunable narrow band limiters \cite{abbas_resonator} can single out the interferer in the attenuation. This, however, still requires prior knowledge of the interfering signal, and limits the instantaneous operating bandwidth of the RF chain. Furthermore, limiters frequently suffer from limited linearity, and recovery times of up to 10--40 ms \cite{gan_robust}.

Another set of technologies to combat interferers in wideband systems is frequency-selective limiters. They typically protect the receiver circuits, while leaving the rest of the channel intact. The innovations in this area include adaptive power limit and tunable frequency. For example, the work in \cite{mansour_self_actuation}, utilizes an intelligent self-actuation of MEMS switches, where a dc bias controls the actuation power level. This provides a frequency-dependent power limiting, which is implemented with a tunable filter. The limited response time of MEMS switches can be resolved with the use of diode-based nonlinear filters \cite{freq_selective_lim_nonlinear_bsf, guyette_sig_track, naglich_freq_selec,hcplr_diode_limiter}. Such power-dependent responses typically have limited power dynamic range (10--20 dB). A wider dynamic range can be achieved with active coupling techniques \cite{wei_selective}, which, similar to \cite{mansour_self_actuation,freq_selective_lim_nonlinear_bsf}, requires prior knowledge of the interfering signal frequency. Spectrum scanning methods in \cite{ak_diplexer,ak_jammer_suppression} are effective in locating the interferer in the band of interest, at the cost of extended response time.

The interference detection method presented in this paper is capable of operating over a multiple octaves, while maintaining sub-$\mu$s response time. The detection method is based on weakly coupling into the incoming signal, and measuring the standing wave pattern of the resulting waveform. The initial concept is presented by the authors in \cite{ak_spiky}. This work significantly expands the on the previous work by analyzing the design and optimistic the tradeoffs. In addition, several new concepts and methods are introduced that result in 1) increasing the instantaneous operating bandwidth by two octaves (from 2--8 GHz to 1--16 GHz), 2) increasing the detected power dynamic range by 20 dB (from 20 dB to 40 dB), 3) halving the response time (from 1 $\mu s$ to 0.5 $\mu s$), and 4) demonstrate the ability to detect and suppress multiple interferers. The interference detection method is implemented and fitted with different types of tunable bandstop filters (BSFs) to demonstrate its versatility. The unmatched measured results represents a feasible solution for wideband receivers operating over multi-octave bands simultaneously.  

%To cover a multi-octave bandwidth, a large number of bandpass filters (BPFs) are required to be engaged simultaneously. Additional limitations include limited power dynamic range and the relatively slow response of MEMS devices. A faster architecture
%
%is important but has many problems 
%many technologies are presented to help
%
%digital methods \cite{gnss_digital_interf_detec,dsp_beamforming,uwb_digital_suppression} mixer-based \cite{mixer_detection}
%
%Limiters attenuate the whole band \cite{mems_self_actuating_limiter,pin_limiter} and sometimes they have linearity issues
%or narrow band with tuning ability, requires prior knowledge of the freqyency \cite{abbas_resonator}
%LNA circuit-based\cite{gan_robust}, 10-40 ms response time, and 
%
%scanning takes forever \cite{ak_diplexer,ak_jammer_suppression}
%
%frequency selective limiters require prpevious knowledge, or slow \cite{mansour_self_actuation,naglich_freq_selec,wei_selective}
%filter bank required prior knowledge and manifold \cite{naglich_freq_selec} bulky and lossy hard to design , hard to scale over multi-octave
%
%In this paper, .. 
%expand on previous one with
%1. 2 more octaves
%2. 2x times faster
%3. 20 dB more power dynamic range
%4. multi interference detection

\begin{figure*}[!t]
\centering
\includegraphics[width=6.5in]{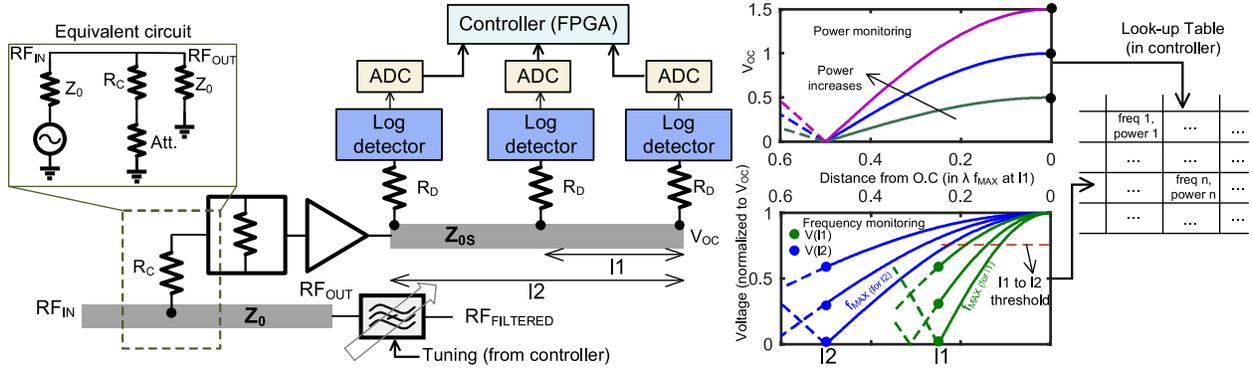}
\caption{The interferer detection concept relies on resistive sampling of the signal (using $R_C$). The amplitude of the sampled signal is then conditioned (with a programmable attenuator and an amplifier) to drive an open stub. The voltage at the open end is proportional to the interferer power. Additional measurement points (at $l1$ and $l2$) are used to measure the standing wave voltage, which carries the interferer frequency information. Multiple point are used to increase the detectable frequency range. A look-up-table is used to relate the voltage readings to interferer power and frequency.}
\label{fig_concept}
\end{figure*}

\section{Detection Theory of Operation}

The initial detection concept is briefly summarized in \cite{ak_spiky}. Some parts below are repeated for completeness. 

\subsection{Detection Concept}
\label{sec_structure}

In the presence of multiple signals in a channel, the waveform is dominated by that of the highest power signal. As a result, the frequency and power of the incoming signal are monitored and used to determine whether the strongest signal is an interferer that should be suppressed. The concept of monitoring the power and frequency of an incoming signal is summarized in Fig. \ref{fig_concept}.

A weak resistive coupling into the received signal path produces a replica of the signal into the monitoring part. A resistor, as opposed to reactive components, is preferred since it results in a frequency-independent coupling. In addition, reactive coupling typically suffers from self-resonance when operating over multi-octave bandwidth. The coupled signal then goes through a programmable attenuator, which maintains the signal to within a manageable level for subsequent stages. This also serves in increasing the power dynamic range of the signal monitoring part. The signal is then amplified to a detectable level.

While, from a noise perspective, the amplifier is expected to be placed before the attenuator, the attenuator is placed first since it typically cannot handle the high power from the amplifier in the case of a strong interferer. This also helps in maintaining a linear gain from the amplifier. 

The signal from the amplifier then drives an open circuit stub. The magnitude of the voltage at the open end is (assuming sufficiently large sensing resistor $R_D$ at the open end)
\begin{equation}
\left|V_{OC} \right| = \sqrt{8P_{STUB}Z_{0S}}= \sqrt{8CL_{Att}G_{PA}P_{in}Z_{0S}}.
\label{eq_voc}
\end{equation}
Where $P_{STUB}$ is the power incident on the open stub, $Z_{0S}$ is its characteristic impedance, $C$ is the power coupling from the $R_C$, $L_{Att}$ is the loss from the attenuator, $G_{PA}$ is the amplifier gain, and $P_{in}$ is the RF input power. The analysis of the circuit model of the resistive coupling (shown in in Fig. \ref{fig_concept}) reveals that the coupling $C$ is 
\begin{equation}
C (dB) = 20 \log \left(  \frac{2Z_0}{2R_C+3Z_0}\right).
\label{eq_cplg}
\end{equation}
Where $Z_0$ is the system impedance ($50~\Omega$ here). From \eqref{eq_voc} and \eqref{eq_cplg}, a smaller coupling resistance provides stronger coupling and larger voltage to detect in the monitoring part of the design. This, however, can deteriorate the insertion loss (S21) and the matching condition (S11). These conditions are described by

\begin{subequations}\label{eq_snp}
\begin{align}
S11 (dB) = 20 \log \left(\frac{Z_0}{2R_C+3Z_0}\right),\label{eq_snp_s11}\\  % there is a negative in S11, but removed since it is in dB
S21 (dB) = 20 \log \left(\frac{2R_C+2Z_0}{2R_C+3Z_0}\right).\label{eq_snp_s21}
\end{align}
\end{subequations}

To properly choose the coupling resistor value, the insertion loss condition is used here since the matching is less sensitive to $R_C$. For example, for a 0.8 dB insertion loss condition, $R_C=210~\Omega$, the coupling $C$=-15 dB, and the matching S11= -21 dB.

The relationship between $V_{OC}$ and $P_{in}$ in \eqref{eq_voc} presents a power monitoring method. This is used in a comparison with a programmable power threshold in order to decide whether to take action against the interferer or not. The minimum and maximum powers are generally dictated by the power detector limits. 

To monitor the frequency, the standing wave pattern is used.

The normalized magnitude of the standing wave pattern is 
\begin{equation}
\left|  \frac{V\left(l\right)}{V_{OC}} \right| =  \left| \cos \left( \frac{\pi}{2} \frac{f}{f_{MAX}} \right)  \right|.
\label{eq_ratio}
\end{equation}
Where $l$ is the length from the open end of the stub, $f$ is the frequency of the strongest signal in the path (potentially the interferer), $f_{MAX}$ is the maximum operating frequency ($c/4l$) where the length $l$ is a quarter wavelength ($\lambda/4$). Beyond this frequency, the relationship between $V(l)$ and the frequency is no longer bijective, and the ratio no longer indicates a unique frequency.

The relationship in \eqref{eq_ratio} dictates the maximum detectable frequency. The minimum detectable frequency, however, is dictated by the resolution of the monitoring system (details in Section \ref{sec_resol}). 

To widen the frequency detection range, a second monitoring node is used at a further distance from the open end ($l2$ in Fig. \ref{fig_concept}). The further monitoring node is used to measure lower frequencies up to its maximum, then the nearer one is used ($l1$). Switching the decision between monitoring $l1$ or $l2$ is based on the value of $l1$ reaching the minimum detectable frequency. 

The relationship in \eqref{eq_ratio} works as the frequency monitoring method. Combined with \eqref{eq_voc}, a two-dimensional calibration table is now formed to identify the frequency and power of the strongest signal in the RF path, which is potentially an interferer as depicted in in Fig. \ref{fig_concept}. 

%More advanced algorithms can be researched to take advantage of both readouts for an even better resolution. 

%
%Structure w frequency extensions
%
%Att then PA has worse noise performance than PA then Att. but Att cannot handle enough power
%
%standing wave equation
%
%power equation

\subsection{Resolution Analysis}
\label{sec_resol}

The ratio in \eqref{eq_ratio} shows small variations at lower frequencies. For example, at $f=0.1 f_{MAX}$ and $f=0.2f_{MAX}$, the values are 0.987 and 0.951, respectively (a difference of 3.7$\%$). While the ratios at $f=0.8 f_{MAX}$ and $f=0.9 f_{MAX}$ are 0.31 and 0.156, respectively (a difference of 15$\%$). In other words, at low frequencies, the monitoring system reaches a point where the ratio no longer indicates frequencies accurately. As a result, it is essential to analyze the resolution of the system to be able to know where to place the secondary sensor, and to quantify the minimum detectable frequency as a specification of the system. 

For the analysis below, $V_{OC}$ is assumed to be held constant at the maximum voltage of the Analog to Digital Converter (ADC). This assumption can be validated by using the programmable attenuator as can be inferred from \eqref{eq_voc}. Furthermore, the equations take into consideration that the employed detectors shown in Fig. \ref{fig_concept} are logarithmic power detectors, to cover a wide sensing dynamic range. 

Using the assumptions above, \eqref{eq_ratio} becomes
\begin{equation}
\log \left(  V\left(l1,2\right) \right) - \log \left(V_{OC} \right) = \log \left( \cos \left( \frac{\pi}{2} \frac{f}{f_{MAX}} \right) \right).
\label{eq_ratio_log}
\end{equation}
The output voltage of the logarithmic power detector, $V_{DET}$, is given by
\begin{equation}
V_{DET}= A\log \left( V\left( l1,2 \right) \right)  +B.
\label{eq_ratio_log2}
\end{equation}
From \eqref{eq_ratio_log} and \eqref{eq_ratio_log2}, we get
\begin{equation}
V_{DET}= A\log \left( \cos \left( \frac{\pi}{2} \frac{f}{f_{MAX}} \right)  \right)  +D.
\label{eq_ratio_log3}
\end{equation}
Where $A$, $B$, and $D$ are constants that can be extracted from the power detector properties and $R_D$. 

\begin{figure}[!t]
\centering
\includegraphics[width=3in]{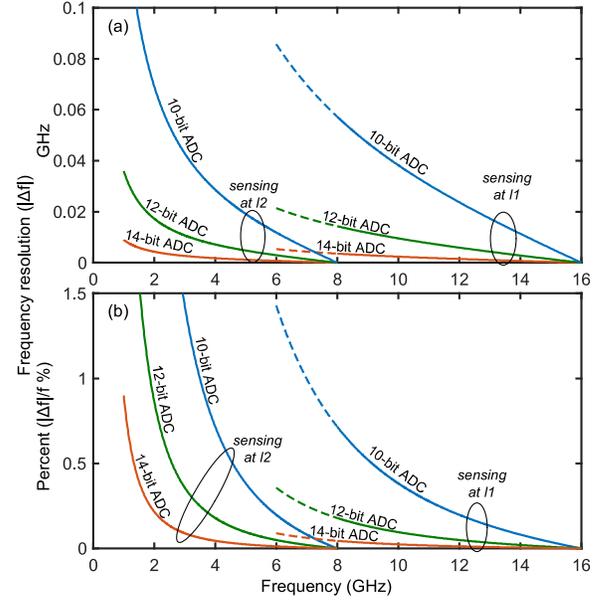}
\caption{The resolution of the sensing method (a) in GHz and (b) in percent of the frequency, as a function of frequency, using various ADC resolution. The percentage numbers are necessary for designing subsequent stages such as BSFs. If the resolution using $l1$ is not accceptable, the systems can switch the sensing to $l2$. The plots are based on A=0.4.}
\label{fig_resolution}
\end{figure}

Then, \eqref{eq_ratio_log3} is differentiated versus frequency, leading to
\begin{equation}
\frac{\partial V_{DET}}{\partial f}= -\frac{A \pi }{2ln(10)f_{MAX}} \tan \left( \frac{\pi}{2} \frac{f}{f_{MAX}}  \right).
\label{eq_dv_df}
\end{equation}
To find the frequency resolution $ \left| \Delta f_{min} \right|$, \eqref{eq_dv_df} is rearranged, and the derivative is approximated as a finite step division, or
\begin{equation}
\left| \Delta f_{min} \right|= \frac{\Delta V_{min}}{\frac{A \pi }{2ln(10)f_{MAX}} \tan \left( \frac{\pi}{2} \frac{f}{f_{MAX}}  \right)}.
\label{eq_res}
\end{equation}
Where $\Delta  V_{min}$ is the minimum change in $V_{DET}$ that is detectable by the ADC. As predicted earlier in this section, it is clear from \eqref{eq_res} that the frequency resolution worsens at low frequencies. As a result, once the minimum detectable frequency change is decided, the further detector (at $l2$) is placed such that its maximum detectable frequency coincides with that minimum acceptable frequency from the nearer detector (at $l1$). These conclusions are illustrated in Fig. \ref{fig_resolution}.

For example, in Fig. \ref{fig_resolution}, if the maximum frequency for $l1$ is 16 GHz, and the maximum accepted $\Delta f$ is 0.25$\%$, then the sensing node $l2$ is placed such that its maximum frequency is at 8 GHz (for 12-bit ADC). Following the 12-bit curve for $l2$ results in a minimum frequency of approximately 3.8 GHz. More sensing nodes can be added to detect even lower frequencies. This, however, increases the length of the sensing stub.

The minimum detectable frequency ($\Delta f$) in \eqref{eq_res} is predicted to be infinitesimally small at $f_{MAX}$. This, however, assumes that the power detector has an unlimited power detection dynamic range. As a result, a finite value of $\Delta f$ is expected in practice.

While a higher number of bits in an ADC can result in a finer frequency resolution, it typically comes at the cost of slower operation. In other words, the response time has to be taken into consideration when deciding the acceptable resolution.

\section{Design Consideration}

\subsection{Power Limitations}

The applications of the detection method include suppressing high-power incoming signals. As a result, a few power limitations are studied below. 

\begin{figure}[!t]
\centering
\includegraphics[width=3in]{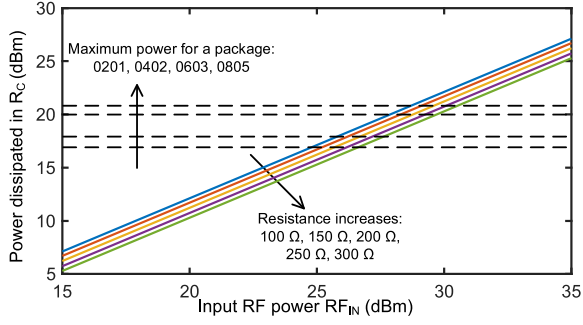}
\caption{The power dissipated in $R_C$ versus the input power at multiple values. The package for $R_C$ dictates the maximum power it can handle, which in turn limits the maximum input average power.}
\label{fig_power_lim}
\end{figure}

The coupling resistor dissipates part of the incoming power. The value of this resistor is typically dictated by the monitoring design as discussed in Section \ref{sec_structure}. As a result, the maximum allowable dissipated power by $R_C$ is dictated by its technology and package. Fig. \ref{fig_power_lim} demonstrated the input power effect on the power dissipated in $R_C$ along with typical power limitations of various packages. This plot can be used to determine the necessary package of $R_C$ for a particular application.

High input power can also saturate the power detector, leading to an erroneous frequency read. This can be concluded from \eqref{eq_ratio}, where a saturated power detector leads to $V_{OC}$ remaining nearly constant, while $V_{l}$ is increasing with power. The programmable attenuator can mitigate this issues by keeping the signal within the linear regime of the detector.

\subsection{Directional Coupler}
\label{sec_cplr}

\begin{figure}[!t]
\centering
\includegraphics[width=3in]{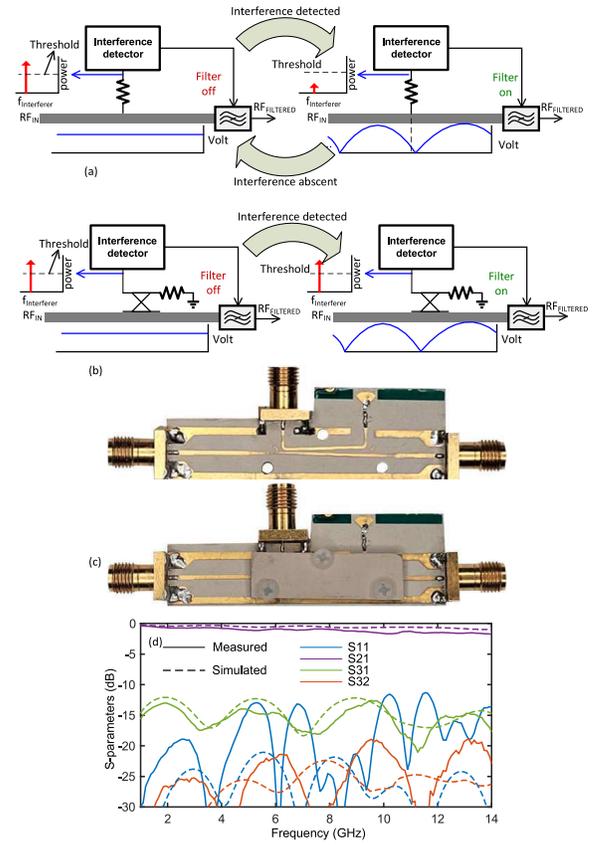}
\caption{(a) If the employed filter is highly reflective, it might cause the detection method to falsely decide that the interferer is no longer present, due to the standing wave pattern created by the filter reflection. (b) A directional coupler can also be used to distinguish the traveling wave from the reflected one. This ensures a stable operation with highly reflective filters at the cost of size. (c) The implemented directional coupler with and without the substrate to cover the striplines. An asymmetric design is used to minimize the footprint. (d) Simulated and measured results of the directional coupler.}
\label{fig_cplr}
\end{figure}

Resistive coupling provides a wideband, compact solution for the detection method. In the presence of a highly reflective filter after it, however, there is a chance that the operation of the detector and the filter becomes unstable. This instability, demonstrated in Fig. \ref{fig_cplr}(a), results from the filter reflecting the interferer, which creates a standing wave between the filter and the preceding circuits including the detection method. If the resistor is at a location where the standing wave has a low value, it will then falsely detect that the interferer is no longer present and the controller can disengage the filter. At which point, the interferer is detected again. 

This instability can be resolved with various methods including reflectionless filters \cite{wei_selective, guyette_absorptive, ak_dual}, and isolators \cite{3d_printed_isolator,isolator_w_diodes}. The bandwidth of these technologies, however, is typically limited to one octave. As a result, a directional coupler is also demonstrated in lieu of $R_C$ in case a highly reflective filter is used. The applied concept is shown in Fig. \ref{fig_cplr}(b).

To minimize the size of the directional coupler, an asymmetric five-section coupled line design is used \cite{cplr_tables}. A picture of the directional coupler is shown in Fig. \ref{fig_cplr}(c). The utilized striplines are sandwiched between two substrates to minimize the mismatch between the even and odd mode propagations in the coupled lines. The radial stub enhances the termination at high frequencies. 

The coupler is fabricated on a 25-mil RO3006 substrate with $\epsilon_r$=6.15. While higher dielectric constants can shrink the overall size, it makes the stripline widths smaller and more sensitive to fabrication tolerances. On the other hand, lower dielectric constants dictate smaller gaps between the coupled lines, which can also be sensitive to tolerances. As a result, an intermediate dielectric constant is used to maintain the widths and gaps reasonably large to be more immune to variations. 

The simulated and measured results show approximately 15 dB of coupling, and better than 1.6 dB insertion loss up to 14 GHz as shown in Fig. \ref{fig_cplr}(d). The directivity is above 6 dB throughout its operating band, which is sufficient for normal operation of the detector.

%measured and simulated 

\section{Implementation and Measurements}

\subsection{Designed Structure}

\begin{figure}[!t]
\centering
\includegraphics[width=3in]{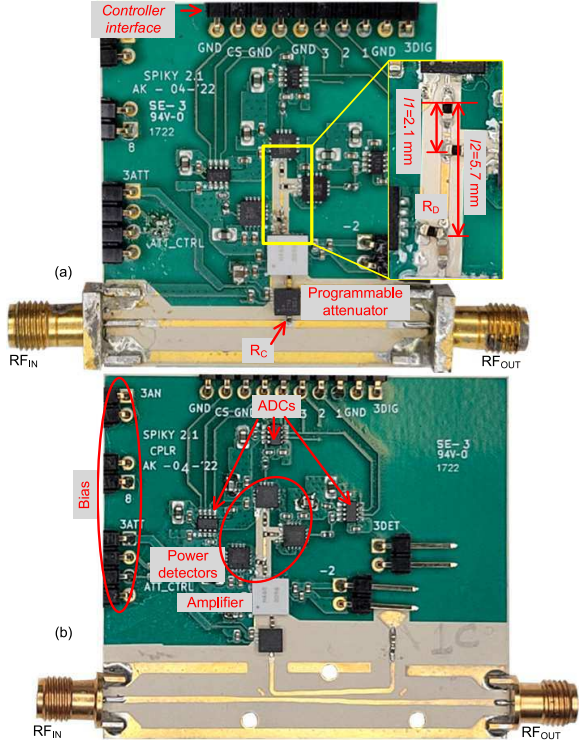}
\caption{The implemented interference detection with (a) resistive coupling, and (b) directional coupler. The directional coupler prevents reflected signals to cause potential false readings. The two designs are identical otherwise.}
\label{fig_photos}
\end{figure}

The interference detection method in Fig. \ref{fig_concept} is prototyped with both coupling methods, resistive and using the directional coupler. The rest of the circuits are similar. The implemented PCBs are shown in Fig. \ref{fig_photos} along with relevant dimensions. Table \ref{table_components} summarizes the components used in this design. The 12-bit ADC is running at 5 MSPS.

The presented prototype consumes 280 mA from a 3.3 V power supply, and 68 mA from an 8 V power supply for the amplifier. Since the power detectors and ADCs are electronic circuits, further power reduction can be achieved with an integrated implementation. 

The detector is designed such that the maximum operating frequency is 16 GHz for the first detector (at $l1$), and 5 GHz for the second detector (at $l2$). The dimensions $l1$ and $l2$ deviate slightly from the theoretical values ($\lambda_g/4$) primarily due to the loading effect from the finite resistors $R_D$. The final values are optimized in simulations. This configuration results in a minimum detectable frequency of 1 GHz. In the prototype with the directional coupler, however, the maximum frequency is limited to 14 GHz as shown in Fig. \ref{fig_cplr}(d). The results from either coupling method are otherwise identical. 

\begin{table}[!t]
\caption{List of components in the interference detection prototype}
\centering
\begin{tabular}{|l|c||l|c|}
\hline
$R_C$ & 220 $\Omega$ & $R_D$ & 180 $\Omega$\\
\hline
Amplifier & HMC460LC5 & Power detector & HMC1094LP3E\\
\hline
Attenuator & RFSA2113 & ADC & LTC2315CTS8\\
\hline
\end{tabular}
\label{table_components}
\end{table}

\subsection{Characterization}

To quantify the performance of the interference detection method, first, its effect on the insertion loss is studied. Since the results from the directional coupler prototype are discussed in Section \ref{sec_cplr}, the resistive coupling is discussed below.

\begin{figure}[!t]
\centering
\includegraphics[width=3in]{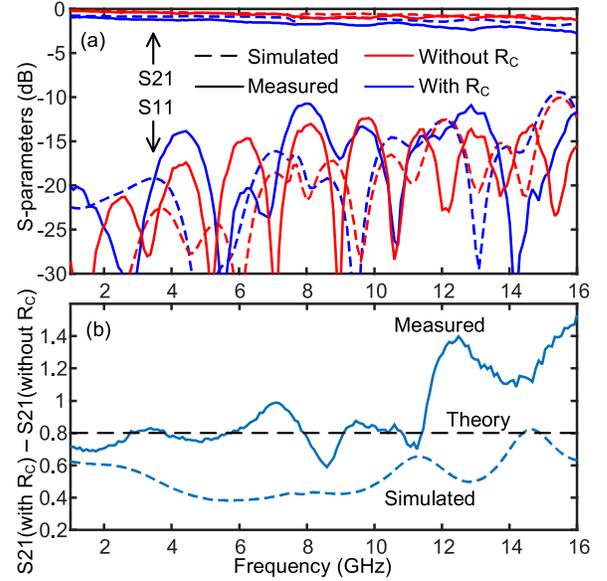}
\caption{(a) Measured and simulated S-parameters of the transmission line in the detection board with and without the coupling resistor $R_C$. The coupling resistor shows low loss and minimal effect on reflection. (b) Zoomed in measured and simulated insertion loss with the resistive coupling compared to the analytical value in \eqref{eq_snp_s21}.}
\label{fig_insertion}
\end{figure}

Fig. \ref{fig_insertion}(a) shows the S-parameters from $RF_{IN}$ to $RF_{OUT}$ with and without the inclusion of the coupling resistor $R_C$. As expected from \eqref{eq_snp_s11} and \eqref{eq_snp_s21} a small attenuation in S21 appears as a result of the coupling, while S11 is nearly unaffected. Fig. \ref{fig_insertion}(b) compares the achieved attenuation with \eqref{eq_snp_s21}. The minor deviation is attributed to the finite reflection of the programmable attenuator, and the frequency-dependent value of the resistors due to package parasitics. 

\begin{figure}[!t]
\centering
\includegraphics[width=3in]{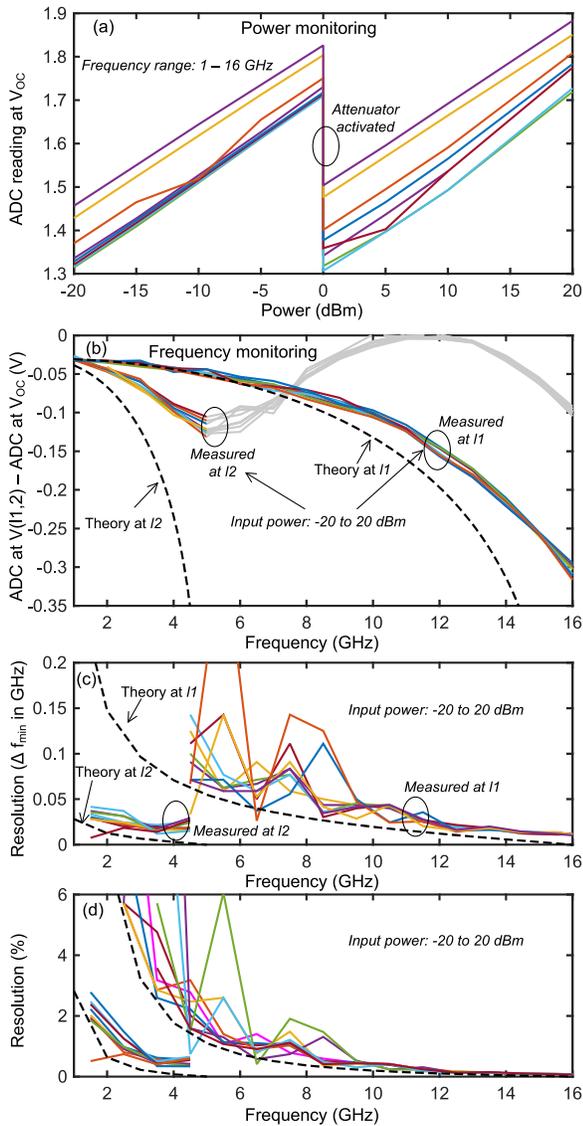}
\caption{(a) Readings from the ADC at the open end of the stub. The voltage is proportional to the input power and is used for power monitoring. The programmable attenuator is activated around 0 dBm to maintain the linear region of the power detector. (b) Analytical and measured values of $V_{DET}$ in \eqref{eq_ratio_log2} at the low frequency point $l2$ and high frequency point $l1$. This represents the frequency monitoring capability. (c) and (d) The frequency resolution of the measured monitoring data in (b) in GHz and in percentage of operating frequency, respectively. All the measurements above are taken from -20 to 20 dBm.}
\label{fig_f_lf_hf}
\end{figure}

The readouts from the ADCs are then investigated versus the input power and frequency (as an interferer). Fig. \ref{fig_f_lf_hf}(a) shows the RF input power versus the reading of the ADC at the open end of the stub. To avoid the saturation of the power detectors, the attenuator is activated at relatively high input power. The power dynamic range is limited by the linear range of the power detectors. The linear (in logarithmic scale) relationship complies with the \eqref{eq_voc}. The frequency-dependent response of the detector does not compromise the frequency/power detection. This is because the detection method also relies on the frequency reading. In other words, the system picks the proper curve in Fig. \ref{fig_f_lf_hf}(a) when determining the power.

For the frequency detection, Fig. \ref{fig_f_lf_hf}(b) shows the normalized power detector readings at $l1$ and $l2$, along with the theoretically predicted values in \eqref{eq_ratio_log3}. It is clear that at lower frequencies, the readings from $l1$ show no significant changes. As a result, the readings from $l2$ are used since they are more accurate. At higher frequencies, on the other hand, readings from $l2$ are no longer bijective as discussed in Sections \ref{sec_structure}, which means they can only be detected from $l1$ readings. The analytical values in Fig. \ref{fig_f_lf_hf}(b) reach a theoretical negative infinite value. This is naturally not achievable in practice since the power detectors saturate at low powers. 

Fig. \ref{fig_f_lf_hf}(a) and Fig. \ref{fig_f_lf_hf}(b) form the calibration look-up-table shown in Fig. \ref{fig_concept}.

The resolution of the presented prototype is shown in Fig. \ref{fig_f_lf_hf}(c), in GHz, and Figs. \ref{fig_f_lf_hf}(d), in percentage, as derived from the results in Fig. \ref{fig_f_lf_hf}(b). The theoretical value in \eqref{eq_res} is also plotted with the 12-bit ADC used in this implementation. Deviations from the theoretical values are contributed to factors such as modeling accuracy and the frequency-dependent response of the components. The results also show that the minimum detectable frequency is below 3$\%$ for most of the frequency range, which is within the capabilities of numerous filter technologies \cite{microstrip_example,ak_vib,saw_example,ak_temp}.

\begin{figure}[!t]
\centering
\includegraphics[width=3.2in]{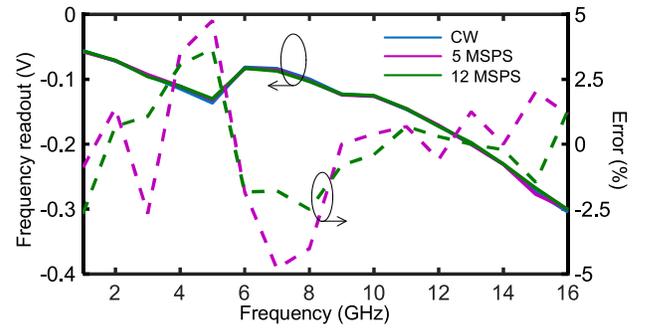}
\caption{Frequency readings (similar to the one in Fig. \ref{fig_f_lf_hf}(b)) for CW signal, in addition to 16QAM signals modulated at 5 and 12 MSPS. The monitoring method shows reasonable robustness versus a modulated signal.}
\label{fig_mod}
\end{figure}

Interferers typically have a finite bandwidth due to modulation. Fig. \ref{fig_mod} shows the reading from the interference detection with continuous wave (CW) signals, in addition to modulated signals with various bandwidths (using 16QAM modulation). The shown measurements are a combination of $l1$ and $l2$, as shown in Fig. \ref{fig_f_lf_hf}(b). The resulting frequency reading shows a weak dependence on the modulation bandwidth, with most of the band being within 2$\%$ error and a maximum deviation of less than 5$\%$. As a result, the detection method is showing a reasonable immunity against interferer bandwidth. While many other combinations of modulation types and bandwidths can be tested, this is beyond the scope of this paper. 

\subsection{Application Measurements}

\begin{figure}[!t]
\centering
\includegraphics[width=3in]{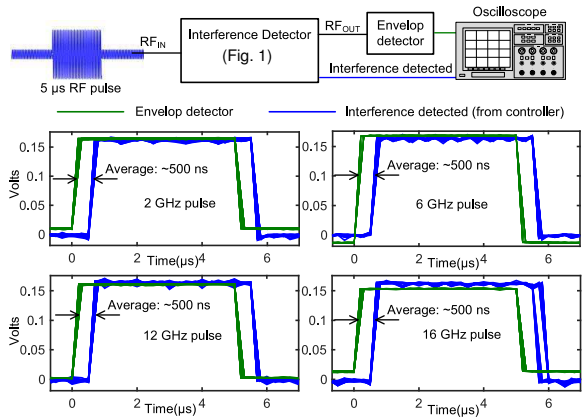}

\begin{picture}(0,0)
\put(-28,147){\fontsize{6pt}{1em}\sffamily(Fig. \ref{fig_concept})}
\end{picture}

\caption{The measurement setup for the response time (top). The measured results show an average of 500 ns response time, independent of interferer frequency. Oscilloscope signals are vertically scaled to clarify the delay time.}
\label{fig_response_time}
\end{figure}

The response time for interference detection is a critical measure to ensure the input energy into a receiver is maintained below the damage level. Furthermore, it is a measure of the downtime of a jammed channel. Fig. \ref{fig_response_time} shows the measurement setup for the response time of the implemented prototype. In summary, an RF pulse is injected into the interference detector, and the output is measured with a wideband envelop detector (ADL6012). The controller in the interference detector also generates a signal indicating the presence of the interferer when it is above a predetermined threshold. The two signals are then measured with an oscilloscope. 

As shown in the plots in Fig. \ref{fig_response_time}, the response time of the interference detector is lagging the RF pulse by 500 ns on average. The delay is primarily due to the ADC (1--2 sample times) and the controller (1 clock time). Faster ADCs can mitigate that if sacrificing the resolution or high power consumption is acceptable.

\begin{figure}[!t]
\centering
\includegraphics[width=2.8in]{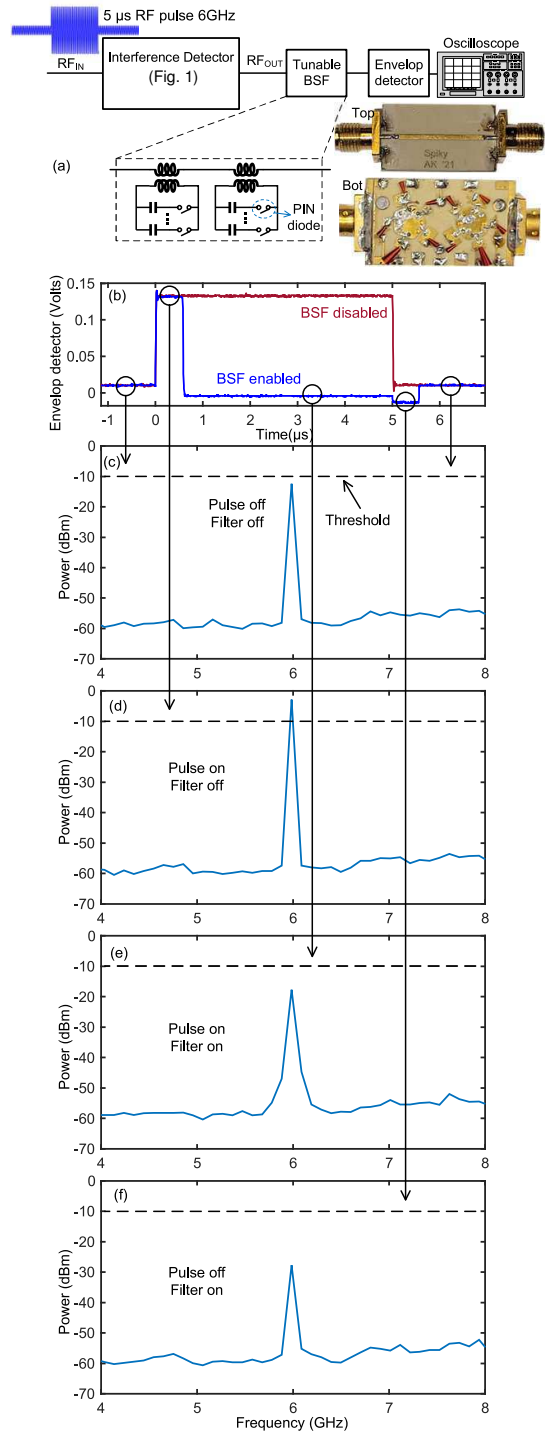}

\begin{picture}(0,0)
\put(-49,523){\fontsize{6pt}{1em}\sffamily(Fig. \ref{fig_concept})}
\end{picture}

\caption{(a) Measurement setup for the response time with a BSF. PIN diode tunable evanescent-mode filters are used to suppress the interferer. (b) Measured envelop of the signal with and without a filter. Initially, the signal has low power and no filter is necessary. This spectrum is shown in (c). Then, a high-power pulse exceeds the threshold as shown in the spectrum in (d). The controller then engages the filter after 500 ns, suppressing the interferer as shown in (e). When the RF pulse ends, the filter remains engaged momentarily, resulting in a spectrum shown in (f). The detector controller then realizes that the pulse is below the threshold, disengaging the filter. The resulting spectrum is also shown in (c)}
\label{fig_response_time_eva}
\end{figure}

To test the response time in a more realistic fashion, the interference detector is fitted with a tunable BSF as shown in the block diagram in Fig. \ref{fig_response_time_eva}(a). The filter is designed with evanescent-mode resonators to achieve a wide tuning range \cite{ak_vib,bsf_liu}. The tuning is achieved with PIN diodes, resulting in a tuning time of 10s ns. The schematic and the photo the implemented filter are also shown in Fig. \ref{fig_response_time_eva}(a). %% XXX AFTER ACCEPTANCE: add more filter references {ak_prox, etc..}

The envelop of the output signal is shown in Fig. \ref{fig_response_time_eva}(b). Compared to the case when the BSF is disabled, the signal is suppressed after the 500 ns delay. When the pulse ends, the filter remains at the pulse frequency for an additional 500 ns before it tunes away from the pulse frequency. The progression of the output spectrum is shown in Fig. \ref{fig_response_time_eva}(c)--(f).

\begin{figure}[!t]
\centering
\includegraphics[width=3in]{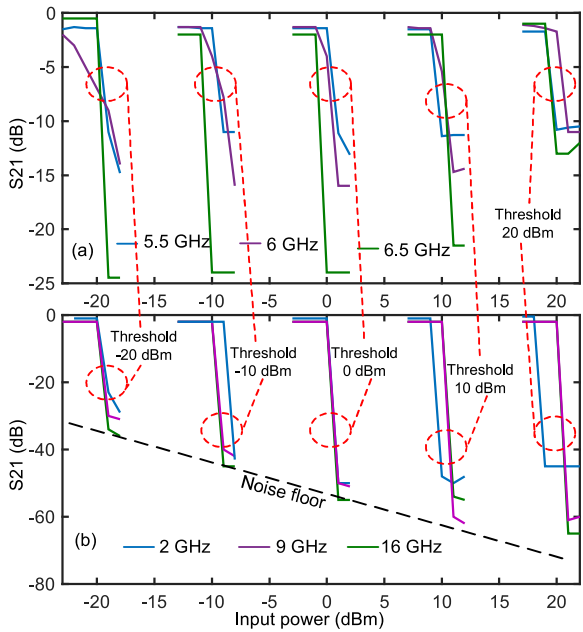}
\caption{Measured signal suppression versus its input power from using the detection method in a programmable limiter. The results are achieved by using (a) evanescent-mode-based filter, and (b) YIG-based filter. The power threshold to engage the filter is programmable between -20 to 20 dBm.}
\label{fig_limiter}
\end{figure}

The interferer detection method can be employed in an automated and programmable power limiter, where it tunes a BSF to the frequency of the interferer. The setup is similar to the one shown in Fig. \ref{fig_response_time_eva}(a). The measurements in Fig. \ref{fig_limiter}(a) is the result using an evanescent-mode-based filter and Fig. \ref{fig_limiter}(b) a commercial YIG-based filter (MLFR-0220 from Micro Lambda Wireless), respectively. This demonstrates the agility of the presented work versus the filtering technique. The results also demonstrate the ability to program the threshold power between -20 to 20 dBm regardless of the operating frequency. The suppression of the evanescent-mode filter weakens at higher powers due to the PIN diodes being no longer in the off state. This can be mitigated with higher reverse bias. 

\begin{figure}[!t]
\centering
\includegraphics[width=3in]{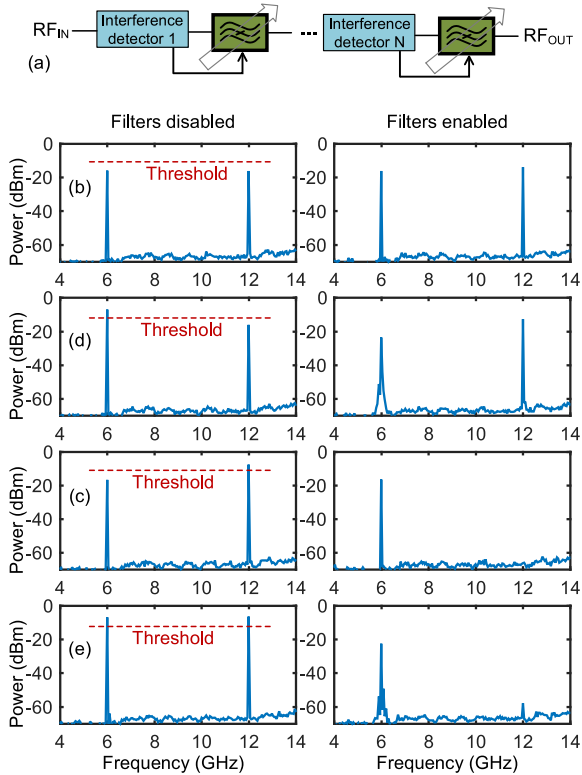}
\caption{(a) Cascading the detectors and filters can suppress multiple interferers. (b)--(e) The output spectra at various power combinations for two interferers at 6 GHz and 12 GHz. The high suppression in (c) and (d) is due to the use of YIG filter at 12 GHz.}
\label{fig_2_interferer}
\end{figure}

In the presence of multiple strong interferers, the presented method can detect them sequentially when cascaded. In such a setup, as shown in Fig.\ref{fig_2_interferer}(a), the first detector/filter combination suppresses the highest power interferer, and subsequent stages follow the power order. This is demonstrated with two interferers as shown in Fig.\ref{fig_2_interferer}(b)--(e) with various combinations. 

\section{Conclusion}
This article presented an interference detection method with a state-of-the-art four-octave frequency detection range, and a response time of 500 ns. The frequency range and resolution of the system are analytically derived for a general implementation. The detection method has the flexibility to interface with a variety of tunable BSFs. This detection method is an enabling technology for wide band receivers, where high-power interferers are a limiting factor.

\section*{Acknowledgments}
This work was sponsored by NSF, award number: 2030257.


\bibliographystyle{IEEEtran}
\bibliography{IEEEabrv,Spiky2p1}

\vspace{11pt}

%\begin{IEEEbiography}[{\includegraphics[width=1in,height=1.25in,clip,keepaspectratio]{M_Abu_Khater}}]{Michael Shell}
%Use $\backslash${\tt{begin\{IEEEbiography\}}} and then for the 1st argument use $\backslash${\tt{includegraphics}} to declare and link the author photo.
%Use the author name as the 3rd argument followed by the biography text.
%\end{IEEEbiography}

\vspace{11pt}

\vfill

\end{document}